\documentclass[aip,jcp,reprint]{revtex4-2}
\usepackage{graphicx}
\usepackage{xcolor}
\usepackage[colorlinks,linkcolor=blue,citecolor=blue,urlcolor=blue]{hyperref}
\usepackage[all]{hypcap}
\usepackage{amssymb}
\usepackage{comment}
\usepackage{xparse}
\usepackage{ifthen}
\usepackage{dcolumn}
\usepackage{siunitx}
\usepackage{booktabs}
\usepackage{ulem}
\usepackage{amsmath} 
\usepackage{bm}

\DeclareSIUnit\angstrom{\text{Å}}
\renewcommand{\thetable}{\arabic{table}}

\NewDocumentCommand{\wf}{O{i}}{\ket{\Phi_{#1}}}
\NewDocumentCommand{\vwf}{s O{i} O{n} O{} O{}}{
  $\chi$_{
    #3}^{#4}
  \IfBooleanTF#1
  {}
  {(\Q_{#5})}
}

\newcommand{\ket}[1]{
  \mathchoice%
  {\left|#1\right\rangle}       
  {|#1\rangle}                  
  {|#1\rangle}                  
  {|#1\rangle}                  
}
\newcommand{\braket}[2]{\left\langle#1\middle|#2\right\rangle}

\begin{document}


\title{Identification of carbon-related color centers in hBN by high-resolution ab initio photoluminescence modeling}

\author{Vytautas Žalandauskas}
\email{vytautas.zalandauskas@fys.uio.no}
\affiliation{Department of Physics/Centre for Materials Science and Nanotechnology, University of Oslo, P.O. Box 1048, Blindern, Oslo N-0316, Norway}

\author{Marek Maciaszek}
\affiliation{State Research Institute Center for Physical Sciences and Technology (FTMC), Vilnius LT--10257, Lithuania}
\affiliation{Faculty of Physics, Warsaw University of Technology, Koszykowa 75, 00-662 Warsaw, Poland}

\author{Marianne Etzelm{\"u}ller Bathen}
\affiliation{Department of Physics/Centre for Materials Science and Nanotechnology, University of Oslo, P.O. Box 1048, Blindern, Oslo N-0316, Norway}

\author{Julius Janušonis}
\affiliation{State Research Institute Center for Physical Sciences and Technology (FTMC), Vilnius LT--10257, Lithuania}

\author{Tadas Paulauskas}
\email{tadas.paulauskas@ftmc.lt}
\affiliation{State Research Institute Center for Physical Sciences and Technology (FTMC), Vilnius LT--10257, Lithuania}

\author{Lukas Razinkovas}
\email{lukas.razinkovas@ftmc.lt}
\affiliation{State Research Institute Center for Physical Sciences and Technology (FTMC), Vilnius LT--10257, Lithuania}

\date{\today}


\begin{abstract}
Identification of optically active defects in hexagonal boron nitride (hBN) remains challenging because different defect candidates can produce similar calculated photoluminescence (PL) lineshapes. Here, we demonstrate that force-constant embedding methodology can be extended to calculate high-resolution, vibrationally resolved PL lineshapes in the polar layered material hBN. Benchmarking against the carbon dimer defect, responsible for the 4.1~eV emission, we show that the methodology accurately reproduces the low-energy acoustic phonons and fine spectral features that are inaccessible when using direct supercell calculations. Applying the methodology to neutral carbon trimers ($\mathrm{C_2 C_B}$ and $\mathrm{C_2 C_N}$), we find that the calculated PL lineshape and excitation energy of $\mathrm{C_2 C_N}$ closely match the low-temperature spectrum of the emitter with a zero-phonon line at 1.54~eV. These results establish high-resolution phonon-sideband modeling as a powerful tool for discriminating among candidate defects in polar materials.
\end{abstract}

\maketitle


\section{Introduction}

Single-photon emission has been observed from color centers in various materials~\cite{Aharonovich_2016,Castelletto}. Among them, hexagonal boron nitride (hBN) is regarded as a particularly attractive platform for single-photon emitters (SPEs) because it combines two key advantages: it hosts bright and photostable SPEs that can operate at room temperature~\cite{Aharonovich_2022}, and its layered van der Waals structure facilitates the transfer and heterogeneous integration of emitter-containing hBN flakes with on-chip photonic devices, including waveguides and optical cavities~\cite{Kim_2019,Froch,Elshaari}. In addition, the use of atomically thin structures, readily realized in hBN, can enhance photon extraction, otherwise limited by total internal reflection in high-refractive-index host materials\cite{vogl2018}.

Despite the many promising aspects of hBN as a host for SPEs, the structural identification of color centers in hBN remains challenging, and typically requires a combination of complementary experimental techniques and first-principles calculations~\cite{Alkauskas_JAP,Freysoldt_RevModPhys}. To date, the microscopic structures of only a small subset of optically active defects in hBN have been established, including the broad near-infrared (NIR) emission band near 1.5~eV, attributed to the spin-active negatively charged boron vacancy~\cite{Gottscholl}, and the ultraviolet (UV) emitter at 4.1~eV, identified as the $\mathrm{C_BC_N}$ carbon dimer defect. The latter assignment was initially proposed based on first-principles calculations~\cite{Mackoit_2019} and subsequently confirmed by combining carbon-isotope substitution and polytype control with ab initio calculations~\cite{Plo}. A similar approach involving polytype control was applied by Iwański et al.~\cite{Iwanski_2024}, again for the dimer, and for the blue quantum emitter (B--center), by Gale et al.~\cite{Gale_2025}. For other emitters, despite extensive research and numerous proposed models, the identification problem has not yet been convincingly resolved~\cite{Cakan,Carbone}.

In hBN, the challenge of microscopic identification is particularly important for the set of NIR~\cite{koperski2020,Konthasinghe,badrtdinov2023,doan2026,paulauskas2026_nir} and red--orange emitters~\cite{Tran2016, Mai,Carbone,paulauskas2026}, which are promising candidates for quantum-photonic applications. The red--orange emitters, with zero-phonon lines (ZPLs) between 1.6 and 2.1~eV, are now known to comprise several spectroscopically distinct families~\cite{Jungwirth_2016,Dietrich_2018,Islam_2024}. Considerable effort has been devoted to identifying the defect structures responsible,
with numerous candidates proposed, including dangling bonds~\cite{Turiansky}, oxygen-related defects~\cite{Li_VBON}, and, most prominently, carbon-related defects~\cite{Auburger,Jara,Maciaszek_2022,Li_trimer,Sajid_2020,Li_CBVBON}. Experimentally, the emitter density was shown to correlate with the carbon concentration in the sample~\cite{Mendelson_2021}, suggesting that at least a subset of these emitters is related to carbon. Among the proposed carbon-related defects, carbon trimers such as $\mathrm{C_2 C_N}$ and $\mathrm{C_2 C_B}$, as well as the carbon-vacancy center $\mathrm{C_B} V_\mathrm{N}$, have received particular attention, with photoluminescence (PL) and photoluminescence excitation (PLE) measurements correlated with ab initio calculations suggesting their possible involvement in red--orange emission~\cite{Tieben,Fischer}.

First-principles modeling of optical spectra enables defect identification through spectral fingerprints linked to their microscopic structure~\cite{Alkauskas_JAP}. However, conventional spectral modeling is limited by the size of computationally accessible supercells, which sample phonons too sparsely to resolve fine spectral detail, and is often insufficient for a convincing structural assignment. Moreover, many defects can produce similar calculated PL spectra, and the identification of the 4.1~eV emitter provides a clear example of this challenge. Specifically, among the proposed candidates for the UV emitter, three yielded calculated spectra in satisfactory agreement with experiment: the carbon dimer~\cite{Linderalv}, the carbon hexamer~\cite{Li_C6}, and the Stone--Wales defect~\cite{Hamdi}. Thus, standard modeling techniques alone may not provide reliable identification, and additional discriminating tests, such as isotope substitution and polytype-dependent spectral shifts~\cite{Plo,Iwanski_2024,Gale_2025}, may be required to resolve subtle differences between candidates. In this study, we apply advanced high-resolution modeling of optical spectra~\cite{Razinkovas_2021} to distinguish among candidate defect structures.

First-principles modeling methods have advanced rapidly in recent years, enabling substantial progress in the simulation of optical spectra from defects. In particular, force-constant embedding approaches, which use data obtained from ab initio-accessible supercells to model the vibrational structure of systems containing more than $10^4$ atoms, have significantly improved the accuracy of optical-spectrum calculations~\cite{Alkauskas_2014, Razinkovas_2021}. This level of modeling now makes it possible to reproduce even subtle features observed in low-temperature PL spectra~\cite{Silkinis_2025}. Complementary machine-learning-based molecular-dynamics approaches have also recently been introduced for modeling temperature-dependent optical lineshapes of color centers~\cite{Linderalv_2025}. So far, however, these various approaches have demonstrated their effectiveness mainly in covalent or weakly polar materials, such as diamond, silicon, and silicon carbide~\cite{Razinkovas_2021,Jin_2021,Jin_2022,silkinis2024,Silkinis_2025,Turiansky_2026,Zalandauskas_2025}. In these materials, the interatomic force constants entering the dynamical matrix are predominantly short-ranged, and longitudinal-optical--transverse-optical (LO--TO) splitting is weak or absent~\cite{Gonze1994, GonzeLee}.

The application of these methods to strongly polar materials has not yet been demonstrated. This issue is particularly relevant for hBN, in which the large electronegativity difference between boron and nitrogen leads to significant polarity. In polar systems, long-range electrostatic dipole--dipole interactions give rise to LO--TO splitting, which can be treated analytically in pristine bulk crystals~\cite{GonzeLee}. However, it remains unclear how these interactions affect defect optical spectra when translational symmetry is broken.

In this study, we demonstrate that high-resolution modeling of low-temperature PL spectra can provide compelling evidence for identifying microscopic defect structures in a polar material, assign the $\mathrm{C_2C_N}$ defect as a source of previously reported experimental data, and reveal clear fingerprints of carbon-carbon bonds in hBN defects. First, we evaluate the performance of our methodology by modeling the PL spectrum of the 4.1~eV emitter in hBN with high accuracy, reproducing subtle spectral features, and strengthening its identification as the carbon dimer $\mathrm{C_B C_N}$~\cite{Mackoit_2019,Plo}. In particular, our approach successfully reproduces the low-energy phonon peak at 10~meV. This agreement validates the accuracy of the embedding methodology for layered materials such as hBN. We then apply high-resolution modeling to the neutral carbon trimers, $\mathrm{C_2C_N}$ and $\mathrm{C_2C_B}$, that were previously proposed as the microscopic origins of red-orange single-photon emission based on the satisfactory agreement between the calculated emission profiles and experimental PL spectra of emitters with ZPLs around 2.1~eV~\cite{Li_trimer, Fischer}. Our results show, however, that the calculated optical spectrum of $\mathrm{C_2 C_N}$ closely matches the low-temperature PL spectrum, as well as the excitation energy, of an emitter in hBN, with a ZPL at 1.54~eV~\cite{badrtdinov2023, Konthasinghe}. This finding indicates that $\mathrm{C_2C_N}$ is unlikely to be responsible for the emitters with ZPLs around 2.1~eV. Our study demonstrates that high-resolution modeling of optical spectra not only deepens the microscopic understanding of the vibrational structure of defects, but also provides a powerful means of discriminating among candidate defect assignments, and shows that, with careful treatment of long-range electrostatic interactions, the approach can be applied to a strongly polar layered material such as hBN.


\section{Results}

\subsection{Bulk hBN lattice dynamics and carbon-dimer benchmark}\label{sec:bulk_hBN_and_dimer}

\begin{figure*}
  \centering
  \includegraphics[width=1\textwidth]{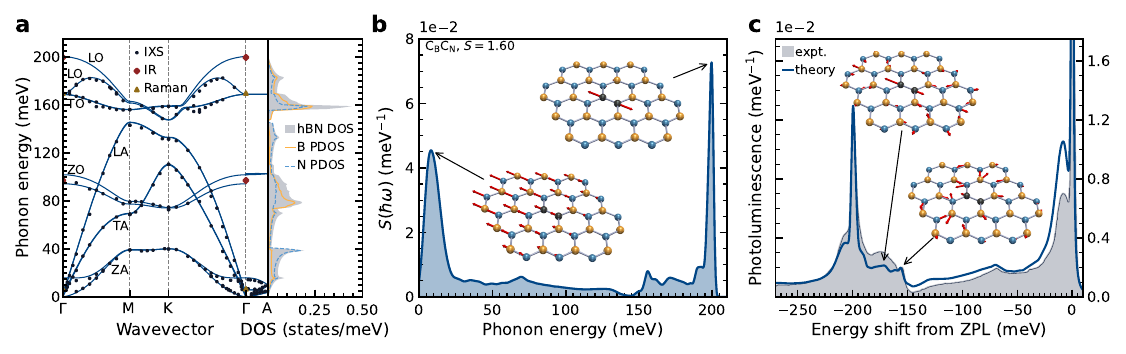}
  \caption{\textbf{Vibrational fingerprints of the carbon dimer defect in hBN.}
  \textbf{a} Left: Phonon dispersion curves of bulk hBN calculated using the r$^2$SCAN functional, compared with experimental inelastic X-ray scattering (IXS) data (black circles) from Ref.~\cite{serrano2007a}. Red diamonds and yellow triangles at the $\Gamma$-point denote infrared (IR)~\cite{geick1966} and Raman~\cite{geick1966,nemanich1981,reich2005} data, respectively. Right: corresponding bulk hBN phonon density of states (DOS) and projected phonon density of states (PDOS).
  \textbf{b} Calculated Huang--Rhys (HR) spectra for the 4.1~eV UV emitter associated with the carbon dimer ($\mathrm{C_B C_N}$) defect. The inset shows the vibrational mode patterns near 10 and 200~meV.
  \textbf{c} Experimental PL spectrum of the 4.1~eV UV emitter from Ref.~\cite{vuong2016}, associated with the carbon dimer defect (gray shaded area, compared with the first-principles PL lineshape (solid blue line). The inset shows the vibrational mode patterns near 156 and 172~meV. Amplitude scaling was optimized for visual clarity.
  \label{fig:C2}}
\end{figure*}

We begin by carrying out density functional theory (DFT) calculations of pristine hBN to establish the reference phonon spectrum for interpreting defect-related signatures in PL phonon sidebands. To describe the lattice vibrations, we employ the r$^2$SCAN meta-GGA functional~\cite{Furness_2020} in combination with the rVV10 nonlocal van der Waals correction~\cite{Peng_2016}. The SCAN family of functionals~\cite{Sun_2015,Furness_2020} has recently been shown to accurately describe the vibrational properties and lineshapes of point defects in several semiconductors, including diamond~\cite{Maciaszek_2023,silkinis2024}, silicon~\cite{Silkinis_2025}, and silicon carbide~\cite{Zalandauskas_2025,younesi2026}. Details of the DFT setup, phonon calculations, and treatment of long-range LO--TO splitting in pristine hBN are described under Methods. 

The calculated phonon dispersion is shown in Fig.~\ref{fig:C2}a together with experimental inelastic X-ray scattering data from Serrano et al.~\cite{serrano2007a}, as well as infrared and Raman data at the $\Gamma$ point~\cite{geick1966,nemanich1981,reich2005}. Bulk hBN has AA$'$ layer stacking and belongs to the $P6_3/mmc$ space group. The phonon spectrum of bulk hBN consists of six mode families, classified as three acoustic (ZA, TA, and LA) and three optical (ZO, TO, and LO) modes. Here, Z denotes out-of-plane polarization, whereas T and L denote in-plane transverse and longitudinal polarizations, respectively. The calculated phonon branches are in good agreement with the experimental data throughout the Brillouin zone, including the high-energy optical modes that dominate the phonon sidebands discussed below. The corresponding phonon density of states (DOS) and projected phonon density of states (PDOS), shown in the right panel of Fig.~\ref{fig:C2}a, contain several pronounced van Hove singularities. The first prominent peak near 40~meV, along with the peak near 130~meV, are primarily associated with nitrogen vibrations. In contrast, the feature near 80~meV is mainly attributed to boron motion.

We now turn to the vibrational structure of the defect--host system. Resolving the phonon sideband in a defect's optical spectrum with high-resolution requires a dense set of normal modes, including long-wavelength acoustic modes, which are not accessible through directly tractable defect-supercell calculations alone. Therefore, we employ the embedding approach developed in previous works~\cite{Alkauskas_2014,Razinkovas_2021}, wherein the bulk Hessian of a pristine supercell is combined with defect force constants obtained from explicit defect-supercell calculations. The agreement between the calculated bulk phonon dispersion and experimental data demonstrates that the computed host force constants serve as a reliable reference for this method. However, the polar structure of hBN complicates the transfer of bulk force constants. In bulk calculations, long-range electric dipole interactions resulting in LO--TO splitting were captured using the non-analytic-term correction of Gonze and Lee~\cite{GonzeLee}. This correction is formulated for a crystal with translational symmetry and cannot be applied directly to an embedded-defect supercell. As a result, only the real-space component of the bulk Hessian is transferable.

We approximately mitigate this issue by calculating the bulk force constants over a large real-space range, ensuring that dipole--dipole interactions are truncated far from the defect. The bulk Hessian is therefore constructed using directly computed force constants from a $16\times16\times3$ repetition of the primitive cell (3\,072 atomic sites). In contrast, the defect force constants are derived from a smaller $7\times7\times2$ defect supercell (392 atomic sites). Despite its size, the pristine supercell is relatively computationally inexpensive because the high symmetry of bulk hBN reduces the number of self-consistent calculations required to obtain force constants to 4, whereas the symmetry-broken defect supercell requires 700. Whether this purely real-space treatment of long-range interactions is sufficient is assessed by the carbon-dimer benchmark that follows. Additional details regarding the embedding procedure and force-constant truncation are provided in the Methods section.

The electron--phonon coupling associated with the optical transition is described within the Huang--Rhys (HR) framework~\cite{huang1950,Alkauskas_2014,Razinkovas_2021}. Coupling to mode $k$ is quantified by the partial HR factor $S_k=\omega_k\Delta Q_k^2/(2\hbar)$, where $\omega_k$ is the vibrational frequency and $\Delta Q_k$ is the transition-induced mass-weighted displacement projected onto the corresponding normal coordinate. The quantity $S_k$ has a direct physical interpretation as the average number of phonons created in mode $k$ during the optical transition. For extended systems, the electron--phonon coupling can be represented by the spectral function, $S(\hbar\omega)=\sum_k S_k \delta(\hbar\omega_k-\hbar\omega)$, which describes the distribution of electron--phonon coupling over vibrational energies~\cite{Alkauskas_2014,Razinkovas_2021}. Additional details regarding the HR formalism, force projection, and optical lineshape calculations are provided in the Methods section.

The performance of this methodology is evaluated by calculating the PL spectrum of the $\mathrm{C_B C_N}$ carbon dimer, which has been identified as the 4.1~eV ultraviolet emitter in hBN~\cite{Mackoit_2019,Plo}. The carbon dimer exhibits a closed-shell singlet ground state, and the optical transition occurs between this state and the excited singlet state. The latter has multireference character and cannot be accurately represented by a single Kohn--Sham determinant~\cite{Mackoit_2019,Maciaszek_2024}. Consequently, the relaxed excited-state geometry is approximated using a mixed-spin $\Delta$-self-consistent-field ($\Delta$SCF method from Ref.~\cite{Gali_2009}) single-determinant configuration, consistent with approaches for carbon-related singlet emitters in hBN~\cite{Mackoit_2019,Maciaszek_2024}. Fig.~\ref{fig:C2}b presents the resulting HR spectrum, calculated using an embedded $40\times40\times4$ supercell comprising 25\,600 atomic sites.

The HR spectrum of the $\mathrm{C_B C_N}$ dimer in Fig.~\ref{fig:C2}b exhibits two pronounced features. The first is a broad peak centered at 10~meV. The effective vibrational motion contributing to this feature, shown in the inset, is obtained by summing the normal modes in the 0--20~meV interval, weighted by their transition-induced displacements $\Delta Q_k$. The resulting coupling is dominated by in-plane acoustic vibrations along the carbon-dimer axis. The second distinct feature at 200~meV arises from coupling to a localized vibrational mode that predominantly involves the stretching motion of two carbon atoms~\cite{Korona_Chojecki}, as illustrated in the corresponding inset.

The computed HR spectrum is then used to obtain the optical lineshape within the zero-temperature approximation using the generating-function approach of Kubo and Lax~\cite{Kubo_1955,Lax_1952}; computational details are provided in the Methods section. In Fig.~\ref{fig:C2}c, the resulting PL lineshape is compared with the low-temperature experimental spectrum from Ref.~\cite{vuong2016}. The theoretical spectrum reproduces all the main features of the phonon sideband. A key success of the embedding methodology is the recovery of the low-energy side peak adjacent to the ZPL. This feature originates from the broad HR contribution near 10~meV and is absent in directly accessible defect-supercell calculations, as shown by the supercell-size convergence analysis in the Supplementary Fig.~1. Although its intensity is slightly overestimated, its presence and spectral position are well reproduced.

In the intermediate-energy region, 40--145~meV, the calculated sideband features align with some characteristic features of the bulk phonon DOS: the 69~meV feature coincides with a small van Hove singularity of TA character, whereas the dip near 147~meV coincides with the gap separating the acoustic and optical manifolds of the bulk phonon DOS. This indicates that this region is dominated by coupling to bulk-like modes. Between 145 and 200~meV, spectral features arise from vibrations of $sp^2$-bonded carbon interacting with hBN bulk modes. It includes the already mentioned localized mode near 200~meV and an additional double-peak structure, comprising a sharp feature near 156~meV and a broader peak near 172~meV. The latter is underestimated compared with the experimental data, possibly due to the approximate treatment of the excited-state geometry. Both features involve carbon motion mixed with host optical phonons, but they differ in character: the sharp feature at 156~meV originates from a quasi-localized vibrational resonance (see Supplementary Fig.~2), whereas the 172~meV feature arises from bulk-like vibrations that couple to the stretching motion of the two carbon atoms, as depicted in the insets of Fig.~\ref{fig:C2}c.

\subsection{Phonon sideband signatures of planar and carbon-related emitters in hBN}

\begin{figure}
  \centering
  \includegraphics[width=0.88\linewidth]{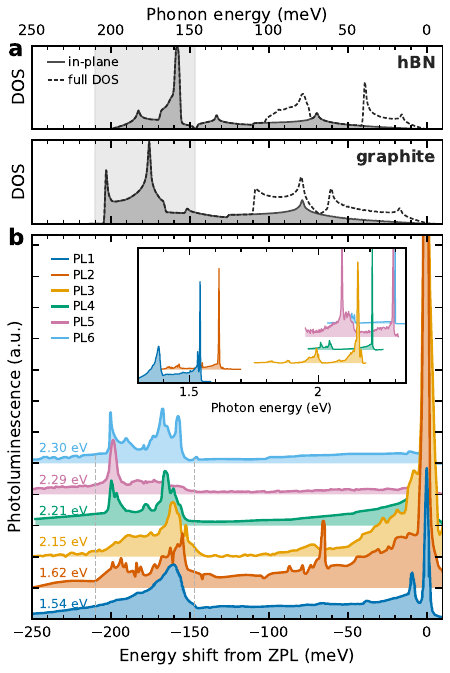}
  \caption{\textbf{Vibrational signatures in low-temperature photoluminescence spectra of hBN single-defect emitters.}
  \textbf{a} Bulk phonon density of states (DOS) of hBN (top) and graphite (bottom), each shown as its in-plane (L$+$T) component (filled) together with the total DOS (dashed); the shaded band marks the optical-sideband window ($-210$ to $-147$~meV) carried down onto panel~\textbf{b}.
  \textbf{b} Phonon sidebands of six single-defect emitters, all measured at cryogenic temperatures and stacked in order of increasing ZPL energy. In the main panel each spectrum is aligned to its ZPL and normalized to the amplitude of its phonon sideband (maximum within the $-210$ to $-140$~meV window), emphasizing the sideband shape; the dashed vertical lines mark the optical-sideband window. The emitters are PL1~\cite{badrtdinov2023} ($1.54$~eV), PL2~\cite{tran2018} ($1.62$~eV), PL3~\cite{Fischer} ($2.15$~eV), PL4~\cite{feldman2019} ($2.21$~eV), PL5~\cite{karankova2026} ($2.29$~eV), and PL6~\cite{fartas2025} ($2.30$~eV).
  \textbf{Inset}: the same spectra on an absolute photon-energy axis, each normalized to its ZPL amplitude, showing the distribution of ZPL energies and the strength of the phonon sideband relative to the ZPL.
}
  \label{fig:expt}
\end{figure}

The carbon-dimer analysis shows that the phonon sideband consists of three types of contributions: in-plane acoustic vibrations at low energies, bulk-like lattice modes that imprint features of the host phonon DOS at intermediate energies, and carbon-related vibrations at high energies that are either localized above the hBN band edge or hybridized with host optical phonons. This picture can be qualitatively extended to other defects that preserve reflection symmetry about the hBN basal plane, including other $sp^2$-bonded carbon complexes. Which host vibrations appear in their sidebands is determined by two factors: (i) symmetry, which restricts the vibrational modes accessible to optical transitions, and (ii) vibrational energy, which determines whether the defect vibrations can hybridize with the lattice modes. First, reflection symmetry restricts linear electron--phonon coupling to modes that are even under reflection through the basal plane, these are the in-plane vibrations (see Supplementary Note~3 for selection rules and the conditions under which they are relaxed). The top panel of Fig.~\ref{fig:expt}a therefore shows the hBN phonon DOS resolved into its in-plane (L+T) component, together with the total DOS.

The second factor is vibrational energy. For planar carbon complexes, the relevant energy scale is set by the $sp^2$ carbon--carbon bond~\cite{bethune1991,ferrari2006}. The lower panel of Fig.~\ref{fig:expt}a shows the in-plane phonon DOS of graphite (see  Supplementary Note~4 for the full dispersion curves and computational details), which serves as a natural reference for these vibrations. Although vibrations in graphite and hBN span a similar energy range, their peak distributions in the DOS differ. Graphite has a dominant TO peak at 176~meV, compared to 158~meV for hBN, while the second LO peak is at 203~meV for graphite, which is above the band edge of the hBN spectrum. This  high-energy peak is characteristic of $sp^2$ carbon--carbon stretching and is responsible for the sharp 200~meV feature observed in the carbon-dimer sideband of Fig.~\ref{fig:C2}c. A single carbon pair embedded in hBN vibrates a few meV below the highest frequency of a fully carbon-bonded network, at the upper edge of the host (i.e., hBN) spectrum where the DOS is negligible. With minimal  hybridization, this vibrational mode is localized almost entirely on the two carbon atoms. In contrast, the lower graphite peak falls well within the hBN optical band resulting in formation of the mode of much stronger hybridization.

Because these two energy ranges overlap, many carbon vibrations of energies below 200~meV in hBN are energetically indistinguishable from host vibrations, resulting in two possible scenarios~\cite{Silkinis_2025}. In the first scenario, the defect induces a resonance, distinguished by enhanced vibrational amplitude near the defect and distributed among multiple host modes within a specific energy window, as observed in the 156~meV feature of the carbon dimer. In the second scenario, carbon atoms participate in host vibrations with amplitudes comparable to those of the surrounding boron and nitrogen atoms. These bulk-like modes also contribute to electron--phonon coupling, but their spectral weight in the sideband depends on two factors: the number of host modes available at a given energy, and the projection of the transition-induced atomic displacement onto each mode. When this projection changes slowly with frequency, the sideband follows the host phonon DOS, as observed throughout the 40--145~meV region of the dimer spectrum. However, when a subset of modes dominates the projection, the sideband deviates from the host phonon DOS. For instance, the second high-energy feature of the dimer appears near 172~meV, close to the graphite TO peak, rather than at the 158~meV maximum of hBN, because the projection favors modes with carbon--carbon stretching character.

With this picture in place, we next examine whether carbon-like vibrational signatures are present in other hBN emitters. Figure~\ref{fig:expt}b displays six representative single-defect emitters whose phonon sidebands were measured at cryogenic temperatures. In each, the primary features fall within the energy range where the optical branches of hBN and graphite overlap. Their ZPLs span the NIR to green spectral region ($1.54$--$2.30$~eV). In the main panel, each spectrum is aligned to its ZPL and normalized to the amplitude of its phonon sideband, supporting direct comparison of sideband shapes between emitters. The inset instead normalizes each spectrum to its ZPL and presents all spectra on a common axis, highlighting the ZPL positions and the relative strength of the phonon sideband.

The two NIR emitters, labeled  PL1~\cite{badrtdinov2023} ($1.54$~eV) and PL2~\cite{tran2018} ($1.62$~eV) in Fig.~\ref{fig:expt}, display qualitatively similar sidebands: each has a pronounced peak approximately $160$~meV below the ZPL, which coincides with the maximum of the hBN phonon DOS and exhibits a weaker feature in the $170$--$210$~meV range, where carbon-related vibrations are expected to contribute. Despite these similarities in sideband shape and ZPL energy, the two emitters differ markedly in the strength of their electron--phonon coupling. As shown in the inset, the sideband of PL2 is minimal relative to its ZPL, whereas PL1 shows a pronounced phonon signature. The sideband of PL3~\cite{Fischer} ($2.15$~eV) closely resembles those of the two NIR emitters, displaying the same dominant $\sim\!160$~meV host feature, even though its ZPL is nearly $0.5$~eV higher in energy.

The remaining three emitters in Fig.~\ref{fig:expt}, PL4~\cite{feldman2019} ($2.21$~eV), PL5~\cite{karankova2026} ($2.29$~eV), and PL6~\cite{fartas2025} ($2.30$~eV), are visible-range centers whose sidebands develop pronounced structure at the upper edge of, and above, the hBN optical band. In addition to the host-related feature near $160$~meV, each exhibits a feature between approximately $185$ and $200$~meV, at or above the upper limit of the bulk hBN phonon spectrum. Although other origins cannot be excluded, this feature is consistent with coupling to a defect-related C--C-stretching vibration, analogous to the $200$~meV localized mode of the carbon dimer. For PL5, this feature is very likely a localized mode: its sideband, a double-peak structure with components near $173$ and $199$~meV (the latter a sharp and intense line), closely resembles the calculated carbon-dimer sideband shown in Fig.~\ref{fig:C2}c.

The emitters depicted in Fig.~\ref{fig:expt} display comparable PL phonon sidebands, each typically marked by two distinct features: a peak appearing around $160$~meV below the ZPL and another at higher phonon energies between $185$ and $200$~meV. We consider the data measured at low temperature since at ambient temperatures thermal broadening would obscure the fine structure, resulting in a double-peak envelope. Likewise, low-resolution spectral simulation would reproduce a similar envelope that does not distinguish between competing defect models. Accurate identification instead requires high-resolution modeling of the vibrational fine structure, compared to low-temperature spectra, in which fine differences in peak positions, lineshapes, and relative intensities become discernible. We provide these comparisons of the complete phonon sideband data of selected defect models in the following section.

\subsection{Carbon trimers and their vibrational fingerprints}

\begin{figure}
  \centering
  \includegraphics[width=1\linewidth]{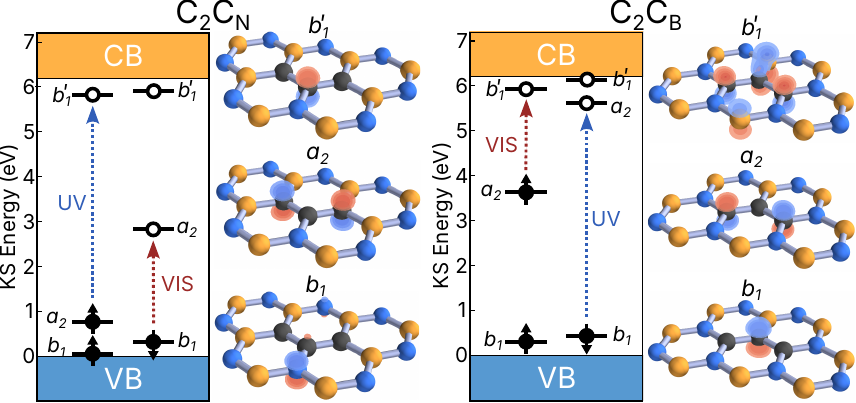}
  \caption{\textbf{Atomic and electronic structure of carbon trimer defects in hBN.} Kohn--Sham single-particle defect levels within the band gap of hBN for the $\mathrm{C_2 C_N}$ (left) and $\mathrm{C_2 C_B}$ (right) configurations, indicating the occupancy of spin-majority (upward arrows) and spin-minority (downward arrows) states, together with optical excitations in the visible (VIS, red dashed arrows) and ultraviolet (UV, blue dashed arrows) spectral ranges. Isosurfaces of the single-particle orbitals associated with the $b_{1}$, $a_{2}$, and $b'_{1}$ defect levels are shown alongside ball-and-stick representations of each defect. Boron, nitrogen, and carbon atoms are shown as yellow, blue, and black spheres, respectively. Blue and red colors denote the positive and negative signs of the orbital wavefunctions.}
  \label{fig:KS_levels}
\end{figure}

Carbon trimers provide a natural test case for vibrational fingerprint analysis. Several previous first-principles studies proposed the $\mathrm{C_2 C_N}$ trimer as the origin of red emission in hBN and reported zero-phonon lines in the vicinity of 1.6~eV~\cite{Li_trimer,Jara,Auburger,badrtdinov2023,tran2018,Golami}, while broader defect surveys also identified both $\mathrm{C_2 C_N}$ and $\mathrm{C_2 C_B}$ as promising carbon-related optical centers~\cite{Maciaszek_2022}. In the following, we focus on these two carbon trimer configurations, and their atomic structures are shown in Fig.~\ref{fig:KS_levels}. We consider the neutral charge state for both trimers, which is also the dominant charge state under the respective growth conditions: previous thermodynamic calculations showed that $\mathrm{C_2 C_N}$ is the relevant trimer under N-poor conditions, whereas $\mathrm{C_2 C_B}$ is favored under N-rich conditions~\cite{Maciaszek_2022}.

The calculated single-particle Kohn--Sham defect levels and orbital isosurfaces for $\mathrm{C_2 C_N}$ and $\mathrm{C_2 C_B}$ are shown in Fig.~\ref{fig:KS_levels}. In both cases, the defect introduces three orbitals in the band gap, labeled $b_{1}$, $a_{2}$, and $b'_{1}$ (according to their irreducible representations) in order of increasing energy, and the corresponding ground state is a spin doublet. A notable feature is that the lowest-lying $b_{1}$ state of $\mathrm{C_2 C_N}$ lies very close to the valence-band edge and has pronounced resonant character, with substantial weight residing on the neighboring nitrogen atom rather than being localized predominantly on the carbon sites. An analogous situation occurs for the highest-lying $b'_{1}$ state of $\mathrm{C_2 C_B}$, which hybridizes strongly with boron-derived states near the conduction-band continuum. Because the degree of hybridization depends sensitively on the energetic proximity to the bulk bands, the detailed orbital character of these states is not fully robust and may vary between electronic configurations; in particular, for $\mathrm{C_2 C_N}$ we find that the character of the $b_{1}$ orbital changes upon imposing the excited-state occupation. This sensitivity limits the usefulness of estimating radiative lifetimes from single-particle transition dipoles, since the corresponding matrix elements depend strongly on the degree of localization of the defect states. Nevertheless, the $\Delta$SCF approach is expected to provide a reasonable approximation to the lowest excited states because it enforces the appropriate orbital occupation and symmetry of the many-electron configuration, although some quantitative discrepancies should be anticipated.

The resulting many-electron states can be understood starting from the single-particle level structure. Because the defect orbitals are nondegenerate, the low-energy many-electron states are expected to be predominantly single-reference in character. In particular, the manifold of doublet excited states can be understood in terms of single-electron promotions between the occupied and unoccupied defect orbitals in either spin channel, while quartet states arise from parallel-spin occupations of the relevant orbitals. For both trimers, however, only one doublet excitation falls into the visible spectral range. For $\mathrm{C_2 C_N}$, this corresponds to promotion of a spin-minority electron from $b_{1}$ to $a_{2}$, whereas for $\mathrm{C_2 C_B}$ it involves promotion of an electron from the mid-gap $a_{2}$ level to $b'_{1}$. These visible excitations are indicated by the red arrows in Fig.~\ref{fig:KS_levels}. All remaining low-lying excitations are expected to lie in the ultraviolet; among them, we also consider the lowest-energy UV transitions, indicated by the blue arrows in Fig.~\ref{fig:KS_levels}.

The calculated ZPL energies for the visible transitions using the HSE06 functional are 1.62 and 1.67~eV for the $\mathrm{C_2 C_N}$ and $\mathrm{C_2 C_B}$ defects, respectively. These values are in good agreement with previous first-principles calculations reported in the literature~\cite{Auburger,Jara,Golami}. Beyond these, the aforementioned UV transitions offer a complementary route for experimentally identifying carbon trimers via correlated visible and UV emission from the same defect center. The corresponding HSE06 ZPL energies are 4.08 and 4.14~eV for the $\mathrm{C_2 C_N}$ and $\mathrm{C_2 C_B}$ defects, respectively.

\begin{figure*}
  \centering
  \includegraphics[width=1\textwidth]{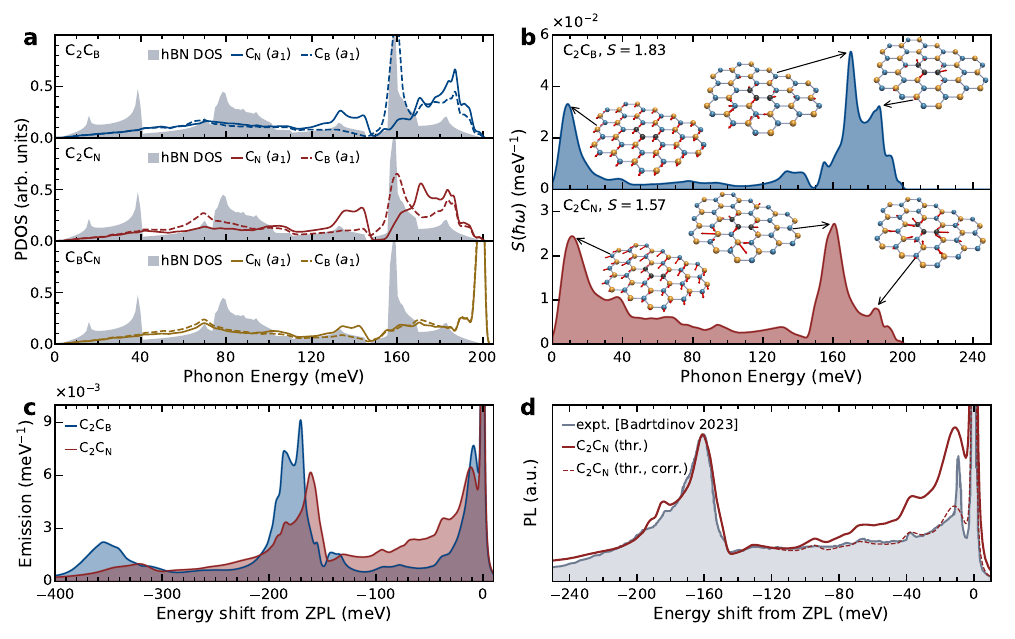}
  \caption{
  \textbf{Theoretical vibrational properties of carbon trimer defects in hBN.}
  \textbf{a} Bulk hBN phonon density of states (DOS) and projected phonon density of states (PDOS) of $\mathrm{C_B}$ (solid lines) and $\mathrm{C_N}$ (dashed lines) atoms in carbon trimer ($\mathrm{C_2 C_B}$ and $\mathrm{C_2 C_N}$) and carbon dimer ($\mathrm{C_B C_N}$) defect complexes.
  \textbf{b} Calculated Huang--Rhys (HR) spectra for the visible optical transition. Bottom: $\mathrm{C_2 C_N}$; top: $\mathrm{C_2 C_B}$. Insets depict the atomic displacements of vibrational modes associated with the peaks at approximately 8, 170, and 186~meV for the $\mathrm{C_2 C_B}$ defect, and 12, 161, and 184~meV for the $\mathrm{C_2 C_N}$ defect. Amplitude scaling was optimized for visual clarity.
  \textbf{c} Calculated vibrationally resolved optical lineshapes for the visible transitions of $\mathrm{C_2 C_N}$ (in red) and $\mathrm{C_2 C_B}$ (in blue) defects.
  \textbf{d} Comparison between the calculated visible-transition phonon sideband of $\mathrm{C_2 C_N}$ (red solid line) and experimental PL spectra from Ref.~\cite{badrtdinov2023}.}
  \label{fig:theory}
\end{figure*}

We next investigate the vibrational properties of the carbon atoms through the projected phonon DOS shown in Fig.~\ref{fig:theory}a. For the $\mathrm{C_2 C_B}$, $\mathrm{C_2 C_N}$, and previously discussed $\mathrm{C_B C_N}$ defects, we show the PDOS projected onto carbon atoms occupying boron sites, $\mathrm{C_B}$, and nitrogen sites, $\mathrm{C_N}$, together with the bulk hBN phonon DOS in gray. Because the electronic states involved in the optical transitions are nondegenerate, only totally symmetric modes participate within the linear Huang--Rhys description. We therefore restrict the carbon projections to modes transforming according to the trivial $a_1$ irreducible representation of the defect point group. Because the $a_1$ modes are symmetric under reflection in the hBN plane, they involve purely in-plane displacements, so the optically relevant carbon vibrations can only hybridize with the longitudinal and transverse phonon branches (Fig.~\ref{fig:C2}a). In contrast, the out-of-plane branches are symmetry-forbidden (see Supplementary Note 3). Below approximately 120~meV, all three defects display similar carbon-projected spectra, reflecting hybridization with acoustic phonons of the host lattice. In particular, the common feature near 69~meV coincides with the van Hove singularity associated with the in-plane transverse-acoustic branch near the $M$ point. Between 130 and 155~meV, the carbon projections exhibit two host-phonon features near 133 and 144~meV, assigned primarily to LA modes, with the $\mathrm{C_N}$ projections carrying greater vibrational weight.

The most pronounced difference between the carbon dimer and the trimers occurs in the optical-phonon region above 150~meV. In the $\mathrm{C_B C_N}$ dimer, the C--C stretching vibration lies above the bulk hBN phonon spectrum and forms a strongly localized mode detached from the host-phonon continuum, giving rise to the pronounced electron--phonon-coupling feature near 200~meV. In the trimers, by contrast, the prominent carbon-related vibrations lie within the hBN optical-phonon continuum and hybridize with the surrounding lattice modes, so no comparably strong, spectrally isolated localized mode is present. The projected spectra of $\mathrm{C_2 C_B}$ and $\mathrm{C_2 C_N}$ are qualitatively similar in this region.

The projected phonon DOS indicates which vibrational modes carry carbon character. However, the electron--phonon coupling that governs the phonon sideband is determined not only by the carbon motion, but also by the overlap of each mode with the structural relaxation that follows the optical transition. The calculated HR spectra for neutral carbon trimer defects are presented in Fig.~\ref{fig:theory}b. Both $\mathrm{C_2 C_B}$ and $\mathrm{C_2 C_N}$ exhibit the same three characteristic electron--phonon coupling regions previously discussed for the $\mathrm{C_B C_N}$ dimer defect: a broad low-energy peak below 20~meV arising from coupling to acoustic phonons, a weaker intermediate-energy region extending from approximately 40 to 145~meV, and a feature-rich region between 145 and 200~meV associated with quasi-localized carbon vibrational resonances with the optical phonons of bulk hBN. For $\mathrm{C_2 C_B}$, the dominant electron--phonon coupling features occur at approximately 8, 170, and 187~meV, whereas the corresponding peaks for $\mathrm{C_2 C_N}$ appear at 12, 161, and 184~meV. The insets in Fig.~\ref{fig:theory}b depict displacements of vibrational modes where the low-energy peak shows long-wavelength in-plane acoustic vibrations. In contrast, the high-energy modes consist primarily of collective in-plane displacements localized on the carbon trimer and its nearest neighbors. The calculated mass-weighted displacements between the ground and excited states, $\Delta Q$, are 0.580 and 0.498~$\mathrm{amu}^{0.5}$\,\r{A} for the $\mathrm{C_2 C_B}$ and $\mathrm{C_2 C_N}$ defects, respectively. The larger structural relaxation of the $\mathrm{C_2 C_B}$ defect upon optical excitation results in stronger electron--phonon coupling, as reflected by the total Huang--Rhys factor, $S_{\mathrm{tot}}$, which is 1.83 for $\mathrm{C_2 C_B}$ and 1.57 for $\mathrm{C_2 C_N}$.

The computed emission lineshapes for the carbon trimer defects are shown in Fig.~\ref{fig:theory}c. Although the overall PSB profiles of the two carbon trimers are qualitatively similar, they exhibit distinct differences in the high phonon energy region. The carbon-related vibrational resonances identified in the HR spectra reappear as pronounced peaks in the PSB at the same energies below the ZPL. Both defects also exhibit a weak shoulder in the 190--200~meV range, originating from quasi-localized vibrations of the carbon trimer, which lie at the top edge of the hBN LO branch, where the phonon frequencies are most sensitive to long-range electrostatic interactions and therefore carry the largest residual uncertainty in our approach. Furthermore, both spectra exhibit broad mound-like features arising from multiphonon replicas, at approximately 355~meV below the ZPL for $\mathrm{C_2 C_B}$, and at 320~meV below the ZPL for $\mathrm{C_2 C_N}$.

With the vibrational fingerprints of both trimers established, we shift our focus to experimental analysis. Among the emitters compiled in Fig.~\ref{fig:expt}, the emitter named PL1, at approximately 800~nm (1.54~eV), exhibits a PSB and ZPL energy that most closely resemble the calculated $\mathrm{C_2 C_N}$ spectrum. Figure~\ref{fig:theory}d presents a comparison between the calculated $\mathrm{C_2 C_N}$ lineshape and the low-temperature PL spectrum of PL1~\cite{badrtdinov2023}. Here, the solid curve is the parameter-free first-principles lineshape, and both spectra are normalized to their dominant high-energy phonon-sideband peak. Although the calculation reproduces the positions of the spectral features throughout the sideband, it overestimates the measured intensity in the low-energy region within about 120~meV of the ZPL. To facilitate a direct comparison of the frequencies of the weak low-energy features, the dashed curve shows the same lineshape multiplied by a smooth monotonic envelope over this region [$f(E)=\exp a(E+120~\mathrm{meV})$], which leaves the high-energy carbon-related features unaffected. Calculations capture the low-energy acoustic-phonon sideband below 20~meV, which gives rise to the characteristic asymmetric ZPL lineshape observed at elevated temperatures. In the intermediate-energy range of 40--145~meV, the PSB features are governed primarily by bulk hBN vibrations and are well reproduced, consistent with the analysis in Sec.~\ref{sec:bulk_hBN_and_dimer}. The prominent electron--phonon coupling peak at approximately 161~meV below the ZPL is also accurately captured and corresponds to a vibrational resonance between carbon-trimer-related modes and bulk hBN optical phonons. This coupling is enhanced by the presence of a van Hove singularity in the bulk phonon density of states. The more localized vibrational resonance is found at approximately 184~meV below the ZPL in the calculations, in good agreement with the experimental feature at around 183~meV. A further high-energy feature appears at approximately 193~meV below the ZPL in the calculated spectrum, which corresponds to a weak and less clearly resolved experimental feature at around 196~meV. This feature lies at the top edge of the hBN optical-phonon band, where frequencies are set by long-range electrostatic interactions whose approximate treatment in our approach may explain the small discrepancy.


\section{Discussion}

Force-constant embedding methodology enables simulation of high-resolution optical spectra of color centers in hBN, a polar layered material. Detailed analysis of electron--phonon interactions and individual sideband features lets us unravel the interplay among features from bulk-like phonons, localized defect vibrations, and quasi-localized vibrational resonances. This approach opens new possibilities to distinguish among candidate atomic structures where conventional supercell calculations cannot.

The method applied to the 4.1~eV carbon dimer emitter demonstrates the validity of the approach. In particular, the embedded calculations reproduce the low-energy acoustic-phonon sideband adjacent to the ZPL, which gives rise to the asymmetric ZPL lineshape at elevated temperatures, the localized carbon--carbon stretching mode near 200~meV, and the carbon-related vibrational resonances at 156 and 172~meV, reproducing the experimental emission lineshape.

Applied to the carbon trimers, the same approach gives a lineshape for $\mathrm{C_2C_N}$ that agrees with the low-temperature PL spectrum of the 1.54~eV emitter~\cite{badrtdinov2023}: the main sideband peaks at 161 and 184~meV coincide with the experimental ones, and the calculated ZPL of 1.62~eV is close to the measured value. For $\mathrm{C_2C_B}$, the Huang--Rhys factor is larger (1.83 versus 1.57), and the high-energy peaks lie at 170 and 187~meV. These differences are large enough to distinguish the two trimers in low-temperature PL spectra.

Our analysis shows that the sideband signatures of $sp^2$-bonded carbon complexes in hBN fall largely within the optical-phonon band of the host, because the vibrational spectra of graphite and hBN span a similar energy range. As a result, most carbon-related vibrations hybridize with host optical phonons and appear in the sideband as quasi-localized resonances rather than as isolated defect modes. Only the highest-frequency C--C stretching vibration can rise above the hBN band edge and form a localized mode, as for the carbon dimer at 200~meV. In the trimers, the corresponding carbon vibrations remain within the band, near 185~meV. Consequently, carbon-related emitters tend to share a similar sideband with two main features near 160~meV and between 185 and 200~meV, and only the fine structure can distinguish them. The earlier assignment of $\mathrm{C_2C_N}$ to the red--orange emitters near 2.1~eV relied on such overall similarity of the sideband~\cite{Li_trimer,Fischer}; when the fine structure is resolved, this trimer is unlikely to account for them.

Regarding the computational methodology itself, the main limitations concern the excited state and long-range electrostatic interactions. Excited-state energies and geometries are obtained with the single-determinant $\Delta$SCF approach. For multireference electronic states, this approximation could affect ZPL energies and the potential energy surfaces that quantify electron--phonon interaction~\cite{Muechler_2022,Ma_2021,Jin_2023}. Approaches beyond this approximation, such as spin-flip time-dependent DFT and the Bethe--Salpeter equation with analytical excited-state forces, have recently been used to compute vibrationally resolved spectra of such states~\cite{Jin_2022,Jin_2026}. The polar nature of hBN also gives rise to long-range electrostatic interactions that complicate the description of lattice vibrations. To partially mitigate these effects, we computed bulk phonons in a very large supercell containing 3072 atomic sites. However, this does not guarantee an accurate description of PSB features near the top of the hBN LO band, where long-range interactions dominate. Extending the treatment of long-range electrostatic interactions from pristine bulk hBN to embedded-defect supercells would remove this limitation.


\section{Methods}

\subsection{Electronic structure calculations}

First-principles calculations were carried out using spin-polarized density functional theory (DFT) as implemented in the \textsc{vasp} package~\cite{Kresse_1993,Kresse_1994,Kresse_1996a}. The projector augmented wave (PAW) method was employed~\cite{Blochl_1994,Kresse_1996b}, and the Brillouin zone was sampled at the $\Gamma$-point. Zero-phonon line (ZPL) energies were computed using the Heyd--Scuseria--Ernzerhof hybrid functional (HSE06) with a screened Fock-exchange fraction of $\alpha = 0.31$~\cite{HSE_2003,HSE_2006}. Long-range dispersion interactions were accounted for using the Grimme-D3 scheme~\cite{Grimme_2006}. ZPL energies and vibrational structure calculations were performed using the r$^2$SCAN meta-GGA functional~\cite{Furness_2020}, which has recently been shown to be suitable for modeling optical and vibrational properties of point defects in hBN and related materials~\cite{Maciaszek_2023,Maciaszek_2024,Chacon_2025,Filippatos_2025,Abbas_2025,Zalandauskas_2025}. Dispersion corrections were included via the rVV10 approach~\cite{Peng_2016}. The plane-wave cutoff energies were set to 500~eV for HSE06 and 600~eV for r$^2$SCAN+rVV10. For HSE06, the electronic self-consistent field (SCF) tolerance and force convergence criteria were $10^{-5}$~eV and \SI{0.01}{\electronvolt\per\angstrom}, respectively. For r$^2$SCAN+rVV10, these criteria were tightened to $10^{-8}$~eV and \SI{0.001}{\electronvolt\per\angstrom}, respectively.

Using the r$^2$SCAN+rVV10 functional, the optimized bulk lattice parameters were $a = 2.502$~\AA\ and $c = 6.587$~\AA, in good agreement with the experimental values of $a = 2.506$~\AA\ and $c = 6.603$~\AA~\cite{Paszkowicz_2002}. The corresponding band gap was 4.83~eV, underestimating the experimental value of 6.08~eV~\cite{Cassabois_2016}. In comparison, HSE06 yielded lattice constants of $a = 2.490$~\AA\ and $c = 6.577$~\AA, and a band gap of 5.95~eV. r$^2$SCAN+rVV10 was used for phonon calculations, both for its substantially lower computational cost relative to hybrid functionals and for better agreement with the experimental lattice constant values.

\subsection{Optical excitation energy calculations}

ZPL energy calculations were performed using a $7\times 7\times 2$ supercell comprising 392 atomic sites. The ZPL values were obtained using the $\Delta$-self-consistent-field ($\Delta$SCF) method~\cite{Gali_2009}, which has been demonstrated to be reliable for carbon-related defects in hBN~\cite{Mackoit_2019,Maciaszek_2024}. The excited-state configurations were modeled by constraining the Kohn--Sham level occupations. For the $\mathrm{C_2 C_N}$ defect, the visible excitation was constructed by promoting a spin-minority electron from the $b_1$ level to the $a_2$ level, whereas the lowest-energy ultraviolet excitation was obtained by promoting an electron from $a_2$ level to the $b_1'$ level in the spin-majority channel. In contrast, for the $\mathrm{C_2 C_B}$ defect, the visible excitation was constructed by promoting an electron from the highest occupied state $a_2$ in the spin-majority channel next unoccupied state $b_1'$ within the same spin channel, while the lowest-energy ultraviolet excitation was modeled by promoting a spin-minority electron from the $b_1$ level to the $a_2$ level.

\subsection{Phonon calculations in the dilute limit}

Vibrational modes for both bulk and defect-containing supercells were obtained via the finite-displacement method, with configurations generated using the \textsc{phonopy} package~\cite{Togo_2015}. Atomic displacements of $\pm 0.01$~\AA\ from the equilibrium geometries were employed. Extrapolation of vibrational modes to the dilute-defect limit followed the embedding methodology of Refs.~\cite{Alkauskas_2014,Razinkovas_2021}. This approach leverages the short-range nature of interatomic forces to construct an effective Hessian matrix for systems containing thousands of atoms. Within this framework, the Hessian matrix elements are determined according to the relative positions of the atoms. For atomic pairs separated by less than a specified bulk cutoff radius, $r_b$, the corresponding elements are taken from the bulk supercell Hessian. For pairs located within a defect cutoff radius, $r_d$, of the defect site, the elements are taken from the defect-containing supercell Hessian. In all remaining cases, the matrix elements are set to zero.

In this work, a large $40\times 40\times 4$ supercell containing 25\,600 atomic sites was constructed by combining force constants extracted from a $16\times 16\times 3$ bulk supercell (3\,072 atomic sites) using a bulk cutoff radius of $r_b = 15.9$~\AA, and a $7\times 7\times 2$ defect supercell (392 atomic sites) using a defect cutoff radius of $r_d = 6$~\AA.

In hBN, the dominant nonanalytic contributions to the interatomic force constants arise from in-plane polarization. Consequently, within the embedding framework, it is desirable to maximize the spatial extent of the cutoff region in the basal plane in order to capture the most significant long-range interactions. In the standard implementation of the embedding methodology, the cutoff region is defined by a spherical radius. However, for layered materials such as hBN, the extent of this sphere is limited along the out-of-plane ($z$) direction by the periodic boundary conditions imposed on the finite supercell. Therefore, we employ an ellipsoidal cutoff region with the out-of-plane axis reduced by a factor of two relative to the in-plane axes. This modification enables the inclusion of a larger number of relevant in-plane interactions while avoiding spurious interactions between periodic images.

To account for the long-range electrostatic interactions arising from the macroscopic electric field generated by collective ionic displacements in hBN, a non-analytical term correction was applied to the Hessian matrix. This correction captures the longitudinal-optical--transverse-optical (LO--TO) phonon mode splitting and was evaluated using the formalism of Gonze and Lee~\cite{GonzeLee}, as implemented in the \textsc{phonopy} package~\cite{Togo_2015}.

\subsection{Optical lineshapes and electron--phonon coupling parameters}

Within the Franck--Condon approximation at absolute zero temperature the normalized optical lineshapes are given by~\cite{Razinkovas_2021}:
\begin{equation}
  \label{eq:lineshape}
  L(\hbar\omega) = C \omega^{\kappa} A(\hbar\omega),
\end{equation}
where $C$ denotes a normalization constant, $A(\hbar\omega)$ is the spectral function of electron--phonon coupling, and the exponent $\kappa$ is equal to 3 for emission and 1 for absorption. The spectral function $A(\hbar\omega)$ is defined by the equation
\begin{equation}
  \label{eq:spectral_function}
  A(\hbar\omega) =
  \sum_{m}
  \left|
    \braket{\chi_{i;0}}{\chi_{f;m}}
  \right|^2
  \delta
  \left(
    E_{\mathrm{ZPL}} \mp (\varepsilon_{fm} - \varepsilon_{f0}) - \hbar\omega
  \right),
\end{equation}
where $\chi_{i;0}$ and $\chi_{f;m}$ are the vibrational wave functions of the initial and final electronic states, respectively. The quantity $E_{\mathrm{ZPL}}$ is the zero-phonon line energy. The energy $\varepsilon_{fm}$ corresponds to the $m$-th vibrational level of the final electronic state, measured relative to the minimum of its potential energy surface. The sign in the argument of the $\delta$ function is minus for emission and plus for absorption processes. The overlap integrals $\braket{\chi_{i;0}}{\chi_{f;m}}$ appearing in Eq.~\eqref{eq:spectral_function} are multidimensional due to the differences between the vibrational structures of the ground and excited electronic states. To facilitate their evaluation, we employ the equal-mode approximation, which assumes that the vibrational mode shapes and frequencies are identical in the initial and final electronic states~\cite{Razinkovas_2021,Markham_1959}.

The spectral function $A(\hbar\omega)$ can be evaluated in the time domain using the generating function approach by Kubo and Lax~\cite{Kubo_1955,Lax_1952}. In this approach, the spectral function is obtained from the generating function $G(t)$ through a Fourier transform:
\begin{equation}
  \label{eq:opt_spectral_func_gen}
  A(\hbar\omega) = \frac{1}{2 \pi} \int_{-\infty}^{\infty}{G(t) e^{-\gamma |t|} e^{- i (E_{\mathrm{ZPL}} / \hbar - \omega) t} \mathrm{d}t},
\end{equation}
where the factor $e^{-\gamma |t|}$ accounts phenomenologically for ZPL linewidth broadening beyond the scope of the present model. The parameter $\gamma$ is adjusted to reproduce the experimental ZPL linewidth and is set to $\gamma = 0.5$~meV in all calculations. Within the equal-mode approximation, the generating function $G(t)$ takes the form
\begin{equation}
  \label{eq:gen_func_with_sum}
  G(t) = \exp \left[ -S_{\mathrm{tot}} + \sum_{k} S_{k} e^{\pm i \omega_{k} t} \right],
\end{equation}
where the plus (minus) sign corresponds to emission (absorption). The total electron--phonon coupling strength is defined as $S_{\mathrm{tot}} = \sum_k S_k$. $S_k$ denotes the partial Huang--Rhys (HR) factor of vibrational mode $k$, and the summation runs over all vibrational modes. $S_k$ represents the average number of phonons created in mode $k$ during an optical transition~\cite{huang1950} and is given by
\begin{equation}
  \label{eq:partial_hr_factor}
  S_k = \frac{\omega_k \Delta Q_{k}^{2}}{2 \hbar},
\end{equation}
where $\Delta Q_k$ denotes the ionic displacement along the $k$-th normal mode induced by the optical transition, defined as the projection of the mass-weighted displacement between the ground- and excited-state equilibrium geometries onto the corresponding normalized phonon eigenvector $\bm{\eta}_{k}$:
\begin{equation}
  \label{eq:delta_q}
  \Delta Q_{k} = \sum_{m\alpha}{\sqrt{M_{m}} \Delta R_{m\alpha} \eta_{k; m\alpha}}.
\end{equation}
Here, $\Delta R_{m\alpha}$ denotes the displacement of atom $m$ along the Cartesian direction $\alpha$, and $M_{m}$ is the corresponding atomic mass.

For extended systems, one must account for a continuum of vibrational frequencies. This is achieved by introducing the spectral density of the electron--phonon coupling, also referred to as the spectral function of electron--phonon coupling:
\begin{equation}
  \label{eq:spectral_function_of_ep_coupling}
  S(\hbar\omega) \equiv \sum_{k} S_{k} \delta(\hbar\omega_{k} - \hbar\omega).
\end{equation}
With this definition, the generating function can be recast in the integral form
\begin{equation}
  \label{eq:gen_func_with_integral}
  G(t) = \exp \left[-S_{\mathrm{tot}} + \int S(\hbar\omega) e^{\pm i \omega t} \,\mathrm{d}\omega \right].
\end{equation}
To obtain a smooth representation of the electron--phonon spectral function, the Dirac delta functions were approximated by Gaussian broadenings with energy-dependent widths, $\sigma$, decreasing linearly from 3.0~meV at low energies to 1.2~meV at the highest phonon energies.

Within the supercell approach, the finite size of the supercells in direct DFT calculations limits the long-wavelength contributions to $\Delta R_{m\alpha}$ in Eq.~\eqref{eq:delta_q}. To recover the dilute-limit relaxation profile, $\Delta Q_k$ is evaluated using the vibrational modes of the large supercell constructed using the embedding methodology. To better capture the long-range component of the transition-induced atomic relaxation, the forces entering Eq.~\eqref{eq:delta_Q_forces} were obtained from $10\times10\times3$ defect supercells containing 1\,200 atomic sites. The displacement along mode $k$ is then given by
\begin{equation}
  \label{eq:delta_Q_forces}
  \Delta Q_{k}
  = \frac{1}{\omega_{k}^{2}} \sum_{m\alpha}\frac{F_{m\alpha}}{\sqrt{M_{m}}} \eta_{k; m\alpha},
\end{equation}
where $F_{m\alpha}$ denotes the force acting on atom $m$ along direction $\alpha$ in the final electronic state evaluated at the equilibrium geometry of the initial state, as obtained from the directly accessible supercell. $\eta_{k; m\alpha}$ is the vibrational eigenmode obtained from the embedded supercell. By combining forces that are already converged at the computationally tractable supercell with the embedding methodology, this approach captures the electron--phonon coupling associated with low-frequency modes and the resulting vibrational resonances.

\section{Acknowledgments}
This work was funded by the Research Council of Lithuania, under Project No.~P-ITP-24-22. The computations were performed on resources provided by UNINETT Sigma2, the National Infrastructure for High Performance Computing and Data Storage in Norway, supercomputer GALAX of the Center for Physical Sciences and Technology, Lithuania, and the High Performance Computing Center “HPC Saulėtekis” in the Faculty of Physics, Vilnius University.

\section{Author contributions}

L.R. and T.P. conceived the idea for this study. V.Ž. and M.M. performed the DFT calculations with supervision from L.R.. V.Ž and L.R. conducted the electron--phonon coupling calculations. V.Ž., J.J. and L.R. analyzed the results. All authors contributed to the writing and editing of the manuscript.

\section{Competing interests}
The authors declare no competing interests.

\bibliography{ref}

\clearpage
\onecolumngrid

\section*{Supplementary Information}

\makeatletter
\setcounter{secnumdepth}{3}
\def\@seccntformat#1{\csname the#1\endcsname\quad} 
\makeatother

\setcounter{section}{0}
\renewcommand{\thesection}{SUPPLEMENTARY NOTE \arabic{section}:}

\setcounter{table}{0}
\renewcommand{\thetable}{\textbf{S\arabic{table}}}
\renewcommand{\theHtable}{S\arabic{table}}
\renewcommand\tablename{\textbf{Supplementary Table}}

\setcounter{figure}{0}
\renewcommand{\thefigure}{\textbf{S\arabic{figure}}}
\renewcommand{\theHfigure}{S\arabic{figure}}
\renewcommand\figurename{\textbf{Supplementary Figure}}

\setcounter{equation}{0}
\renewcommand{\theequation}{S\arabic{equation}}
\renewcommand{\theHequation}{S\arabic{equation}}

\section{Supercell size convergence using embedding methodology}

\begin{figure} 
    \centering 
    \includegraphics[width=1.0\linewidth]{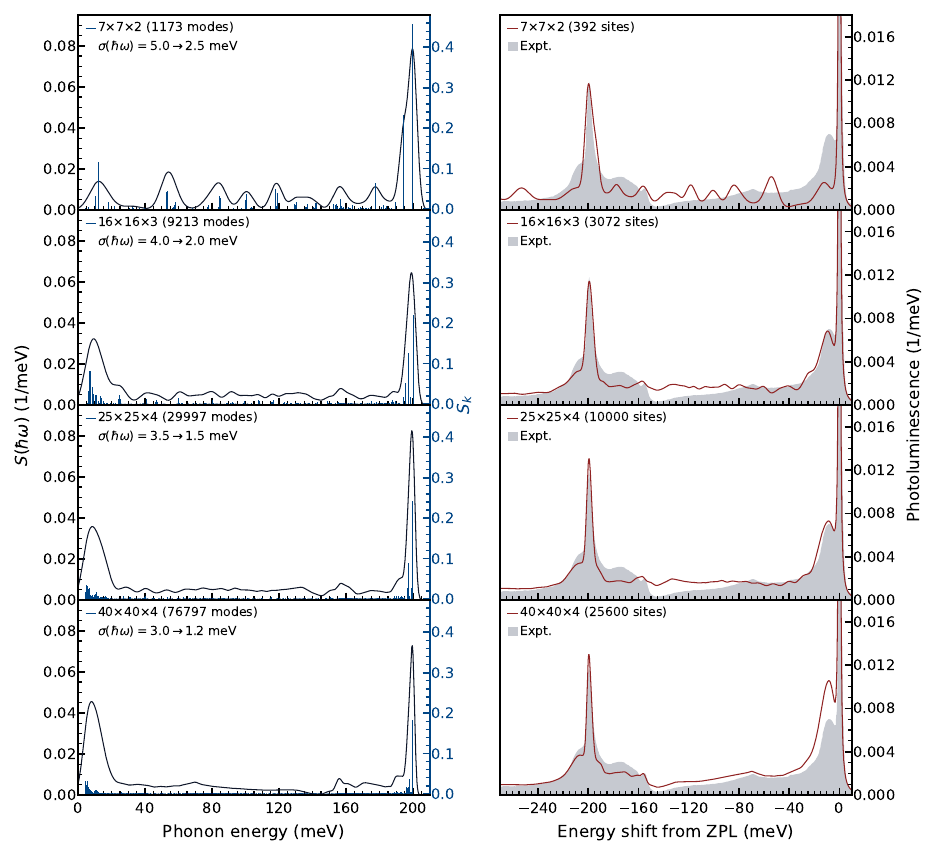} \caption{Convergence of the electron--phonon spectral density $S(\hbar\omega)$ (solid black lines), partial Huang--Rhys factors $S_k$ (vertical blue bars), and theoretical photoluminescence lineshapes (solid red lines) with increasing supercell size. Supercells range from $7\times7\times2$ (392 atomic sites) to $40\times40\times4$ (25\,600 atomic sites). The lineshape for the $7\times7\times2$ supercell was calculated without the embedding methodology. The gray shaded area shows the experimental spectrum from Ref.~\cite{vuong2016}.}
    \label{fig:sm_C2_embedding_conv}
\end{figure}

To assess the convergence of the electron--phonon coupling and the corresponding photoluminescence lineshapes obtained using the embedding methodology described in the main text and in greater detail in Refs.~\cite{Alkauskas_2014,Razinkovas_2021}, embedded supercells ranging from $7\times7\times2$ (392 atomic sites) to $40\times40\times4$ (25\,600 atomic sites) were considered, with intermediate sizes of $16\times16\times3$ (3\,072 atomic sites) and $25\times25\times4$ (10\,000 atomic sites). Left column of Fig.~\ref{fig:sm_C2_embedding_conv} shows the convergence of the electron--phonon spectral density $S(\hbar\omega)$ (solid black lines) and the partial Huang--Rhys factors $S_k$ (vertical blue bars), while right column shows corresponding calculated photoluminescence lineshapes (solid red lines), as a function of supercell size. The calculated lineshapes are compared with the experimental spectrum from Ref.~\cite{vuong2016}.

To obtain a smooth representation of the electron--phonon spectral density, the Dirac delta functions were approximated by Gaussian broadenings with energy-dependent widths, $\sigma(\hbar\omega)$, decreasing linearly from 5.0 to 2.5~meV over the phonon-energy range for the $7\times7\times2$ supercell, from 4.0 to 2.0~meV for the $16\times16\times3$ supercell, from 3.5 to 1.5~meV for the $25\times25\times4$ supercell, and from 3.0 to 1.2~meV for the $40\times40\times4$ supercell. The lineshape for the $7\times7\times2$ supercell was calculated without the embedding methodology. For the larger supercells, the force constants were constructed using the embedding approach from a $16\times16\times3$ bulk supercell (3\,072 atomic sites) with a bulk cutoff radius of $r_b = 15.9$~\AA\ and a $7\times7\times2$ defect supercell (392 atomic sites) with a defect cutoff radius of $r_d = 6$~\AA.

The convergence tests demonstrate that the embedding methodology expands the vibrational basis from 1\,173 vibrational modes in the $7\times7\times2$ supercell to 76\,797 modes in the $40\times40\times4$ supercell, corresponding to an increase by a factor of approximately 65. This substantially denser sampling of the vibrational spectrum enables the use of smaller Gaussian broadening widths $\sigma$, resulting in improved spectral resolution of the electron--phonon spectral density and photoluminescence lineshape. Furthermore, the embedding methodology captures long-wavelength, low-energy vibrational modes that are absent in conventional finite supercell calculations because their wavelengths exceed the dimensions of the explicitly calculated supercell.

\section{Vibrational mode localization ratios}

The vibrational character of phonon modes can be assessed by quantifying their degree of spatial localization~\cite{Bell_1970}. Bulk-like modes are delocalized throughout the crystal and exhibit displacement patterns resembling plane-wave vibrations. In contrast, localized modes are confined primarily to the defect region and often appear at frequencies outside the phonon spectrum of the pristine host material. An intermediate case can be described as quasi-local modes~\cite{Alkauskas_2014}, which correspond to defect-induced vibrational resonances that are embedded within the continuum of bulk phonons. These resonances arise from a redistribution of vibrational character among a group of modes with energies lying within the bulk phonon spectrum. To quantify the spatial localization of vibrational modes, we use the inverse participation ratio (IPR), defined for a phonon mode $k$ as~\cite{Bell_1970,Alkauskas_2014}
\begin{equation}
  \label{eq:ipr}
  \mathrm{IPR}_{k}
    = \frac{1}{\sum_{m} \boldsymbol{\eta}_{k;m}^{4}},
\end{equation}
where $\boldsymbol{\eta}_{k;m}$ denotes the normalized, three-dimensional mass-weighted displacement vector of atom $m$ in mode $k$. The fourth power of the displacement amplitude is defined as
\begin{equation}
  \boldsymbol{\eta}_{k;m}^{4}
    \equiv \left( \sum_{\alpha} \eta_{k;m\alpha}^{2} \right)^{2},
\end{equation}
where $\eta_{k;m\alpha}$ represents the mass-weighted displacement of atom $m$ along the Cartesian direction $\alpha$. The IPR can be interpreted as an effective number of atoms participating in a given vibrational mode. A value of $\mathrm{IPR}_{k}=1$ corresponds to a mode fully localized on a single atom, whereas $\mathrm{IPR}_{k}=N$ indicates a completely delocalized mode in which all $N$ atoms of the supercell contribute equally. For intermediate cases, where approximately $P<N$ atoms contribute significantly to the mode, the IPR approaches $\mathrm{IPR}_{k}\approx P$. For comparison of localization properties across systems with varying numbers of atoms, it is convenient to introduce the normalized localization ratio $\beta_{k}$~\cite{Alkauskas_2014}, defined as
\begin{equation}
  \label{eq:beta}
  \beta_{k} = \frac{N}{\mathrm{IPR}_{k}}.
\end{equation}
Larger $\beta_{k}$ values correspond to stronger localization of phonon mode $k$. The localization ratio is therefore particularly useful for comparing defect-induced vibrational features across different supercell sizes.

\begin{figure} 
    \centering 
    \includegraphics[width=0.7\linewidth]{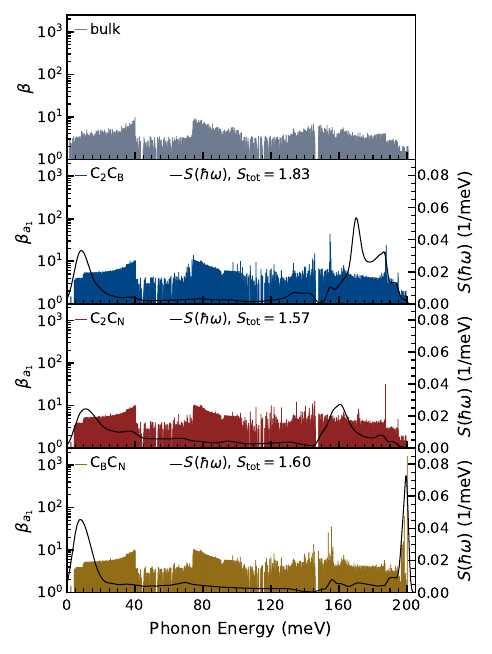} \caption{Vibrational localization ratios $\beta$ (vertical bars) and corresponding electron--phonon spectral functions $S(\hbar\omega)$ (solid black lines) for bulk hBN and carbon-related defects, including the carbon trimer defects $\mathrm{C_2 C_B}$ and $\mathrm{C_2 C_N}$ and the carbon dimer defect $\mathrm{C_B C_N}$.}
    \label{fig:sm_mode_localization}
\end{figure}

Figure~\ref{fig:sm_mode_localization} shows the calculated localization ratios for vibrational modes in bulk hBN ($\beta$, vertical bars) together with the localization ratios of the $A_1$-symmetry vibrational modes ($\beta_{A_1}$) of the ground states of the carbon trimer defects $\mathrm{C_2 C_B}$ and $\mathrm{C_2 C_N}$ and the carbon dimer defect $\mathrm{C_B C_N}$. The corresponding electron--phonon spectral functions $S(\hbar\omega)$ are overlaid as solid black lines. Because the electronic ground states of these defects are nondegenerate, the first-order electron--phonon interaction is restricted by the Wigner--Eckart theorem to totally symmetric vibrational modes. Consequently, only phonon modes of $A_1$ symmetry couple to the optical transition and contribute to the electron--phonon interaction. 

Bulk-like vibrational modes, characterized by relatively small localization ratios ($\beta<10$), are spatially delocalized and retain the collective vibrational character of the pristine hBN lattice and are present in defect-containing systems. Among the carbon-related defects considered here, vibrational modes with high localization ratios ($\beta \geq 100$) are found only for the carbon dimer defect. These correspond to a highly localized vibrational mode dominated by the relative motion of the two carbon atoms, resulting in a mode that is strongly confined to the defect region. Localization ratios of approximately $10 \lesssim \beta \lesssim 100$ are associated with quasi-local vibrational resonances. These modes remain embedded within the continuum of bulk phonons but possess enhanced vibrational amplitudes near the defect. Their signatures are evident in the electron--phonon spectral functions $S(\hbar\omega)$, where several pronounced resonances coincide with peaks in the localization ratio. 

High localization ratio alone does not imply strong electron--phonon coupling. For a vibrational mode to contribute significantly to the electron--phonon interaction, its displacement must also exhibit a substantial projection onto the atomic relaxation profile $\Delta\mathbf{R}$ between the ground and excited states.

\section{Selection rules for vibrations}

Every atomic layer of bulk hBN ($P6_3/mmc$, AA$'$ stacking) is a mirror plane of the crystal, and the site symmetry of the lattice sites and hexagon centers is $D_{3h}$. A defect confined to a single layer therefore retains the basal reflection $\sigma_h$, and its point group is one of $D_{3h}$, $C_{3h}$, $C_{2v}$, or $C_s$. The carbon dimer and trimers belong to $C_{2v}$, with the $C_2$ axis in the layer. We choose $z$ along $C_2$ and $x$ normal to the layer, so that the $\sigma_h$-odd irreducible representations are $B_1$ and $A_2$ (the $\pi$-type defect orbitals $b_1$, $a_2$ of the main text), and displacements of an atom lying in the layer transform as $A_1 \oplus B_2$ (in-plane) and $B_1$ (out-of-plane). The classification for all planar groups is given in Supplementary Table~\ref{tab:sm_selection}.

The linear electron--phonon interaction is characterized by the vibronic coupling constants~\cite{bersuker1989,Razinkovas_2021,Razinkovas_thesis}
\begin{equation}
  V^{ij}_{k} = \langle \Psi_i | \partial U/\partial Q_k | \Psi_j \rangle ,
  \label{eq:sm_vibronic}
\end{equation}
where $\Psi_i$ and $\Psi_j$ are electronic states transforming as the irreducible representations $\Gamma_i$ and $\Gamma_j$,  $U$ is the electron--nuclear potential, and $Q_k$ is a normal coordinate transforming as $\Gamma_k$. For a nondegenerate state, $i=j$ and $V^{ii}_{k}$ is the force along $Q_k$ that drives the lattice relaxation. When $\Psi_i$ and $\Psi_j$ belong to the same degenerate manifold, $V^{ij}_{k}$ are the coupling constants of the Jahn--Teller problem. By the Wigner--Eckart theorem~\cite{tinkham2003,elliott1979,cornwell1997}, $V^{ij}_{k}$ vanishes unless $\Gamma_k \subset \Gamma_i \otimes \Gamma_j$. For a nondegenerate state $\Gamma_i \otimes \Gamma_i = A_1$~\cite{tinkham2003}, so the force exerted by the excited state at the ground-state geometry, and with it the relaxation $\Delta\mathbf{R}$ and all $\Delta Q_k$, are totally symmetric and only $a_1$ modes acquire partial Huang--Rhys factors. Since $C_{2v}$ has no degenerate representations, this covers all transitions considered here. Because $a_1$ is even under $\sigma_h$, every linearly coupled mode is $\sigma_h$-even. For atoms in the defect layer this means purely in-plane motion. The sideband is built from the in-plane (LA, TA, LO, TO) branches of hBN; the ZA and ZO branches do not contribute at first order. Atoms of the adjacent layers are not on the mirror plane and may move out of plane in mirror-related pairs within $a_1$ modes.

Degenerate ($E$) states require a threefold axis and, for planar defects, occur only in $D_{3h}$ and $C_{3h}$. For an $E$ state the Wigner--Eckart theorem admits all modes in the symmetric square $[E\otimes E]$, whose non-totally-symmetric part defines the Jahn--Teller-active modes~\cite{bersuker1989,davies1981}. In both groups $[E'\otimes E']$ and $[E''\otimes E'']$ contain only $E'$ besides the totally symmetric representation (Supplementary Table~\ref{tab:sm_selection}), so the Jahn--Teller-active $e'$ modes are again in-plane. The dynamical $E\otimes e$ problem then has to be solved beyond the Huang--Rhys approximation~\cite{Razinkovas_2021}, but the modes remain in-plane. Out-of-plane modes enter only through pseudo-Jahn--Teller coupling between states of opposite $\sigma_h$ parity~\cite{bersuker1989}, which is a second-order effect.

The rules are relaxed when $\sigma_h$ is lost. This can happen through the defect itself (nonplanar defects, complexes spanning several layers, pseudo-Jahn--Teller buckling in one of the electronic states) or through its environment (rhombohedral stacking, stacking faults, and, weakly, a substrate). Second-order coupling and Herzberg--Teller intensity borrowing~\cite{azumi1977}, which lie outside the linear Condon approximation, can also involve $\sigma_h$-odd modes, but are weak for the dipole-allowed, in-plane polarized transitions studied here.

We compute the partial Huang--Rhys factors in the embedded supercells for all modes without imposing these rules. The $\sigma_h$-odd ($b_1$, $a_2$) modes carry less than $10^{-12}$ of the total Huang--Rhys factor for all three defects, and the $\sigma_v$-odd $b_2$ modes less than $0.5\%$.

\begin{table}
  \centering
  \caption{Point groups of defects in hBN that retain the basal mirror plane $\sigma_h$, with their irreducible representations sorted by $\sigma_h$ parity. For an atom in the basal plane, in-plane displacements span only $\sigma_h$-even and out-of-plane displacements only $\sigma_h$-odd irreducible representations. The last two columns list the modes that couple linearly to a nondegenerate state and, where $E$ states exist, the Jahn--Teller-active modes from the symmetric square $[E\otimes E]$. The lower part lists the trigonal groups reached when $\sigma_h$ is broken; their $e$ modes contain both in-plane and out-of-plane displacements.}
  \label{tab:sm_selection}
  \begin{ruledtabular}
  \begin{tabular}{lllll}
    Group & $\sigma_h$-even & $\sigma_h$-odd & Nondegenerate state & $E$ state (Jahn--Teller) \\
    \colrule
    $D_{3h}$ & $A_1'$, $A_2'$, $E'$ & $A_1''$, $A_2''$, $E''$ & $a_1'$ & $[E'\otimes E'] = [E''\otimes E''] = A_1' \oplus E'$: $e'$ \\
    $C_{3h}$ & $A'$, $E'$ & $A''$, $E''$ & $a'$ & $[E'\otimes E'] = [E''\otimes E''] = A' \oplus E'$: $e'$ \\
    $C_{2v}$ & $A_1$, $B_2$ & $A_2$, $B_1$ & $a_1$ & --- \\
    $C_{s}$  & $A'$ & $A''$ & $a'$ & --- \\
    \colrule
    $C_{3v}$, $D_3$ & \multicolumn{2}{l}{no $\sigma_h$} & $a_1$ & $[E\otimes E] = A_1 \oplus E$: $e$ \\
    $C_{3}$  & \multicolumn{2}{l}{no $\sigma_h$} & $a$   & $[E\otimes E] = A \oplus E$: $e$ \\
  \end{tabular}
  \end{ruledtabular}
\end{table}

\section{Graphite calculations}

First-principles calculations of bulk graphite were performed within density functional theory (DFT) as implemented in the \textsc{vasp} package~\cite{Kresse_1993,Kresse_1994,Kresse_1996a}. The projector augmented-wave (PAW) method was employed~\cite{Blochl_1994,Kresse_1996b}, and the Brillouin zone was sampled at the $\Gamma$ point. The r$^{2}$SCAN functional~\cite{Furness_2020} was used, with van der Waals interactions accounted for using the rVV10 approach with $b=15.7$~\cite{Peng_2016}. A plane-wave energy cutoff of 500~eV was employed, and the electronic SCF and ionic force convergence criteria were set to $10^{-8}$~eV and \SI{0.001}{\electronvolt\per\angstrom}, respectively. The optimized lattice parameters, $a=2.458$~\AA\ and $c=6.719$~\AA, are in good agreement with the experimental values of $a=2.459$~\AA\ and $c=6.672$~\AA~\cite{baskin1955}.

The bulk graphite phonon dispersion and phonon density of states (DOS) were calculated using the finite-displacement method, with displaced configurations generated using \textsc{PHONOPY}~\cite{Togo_2015}. Atomic displacements of $\pm 0.01$~\AA\ from the equilibrium positions were employed. A $7\times7\times2$ supercell containing 392 atoms was used for the phonon calculations.

\begin{figure}
    \centering 
    \includegraphics[width=0.8\linewidth]{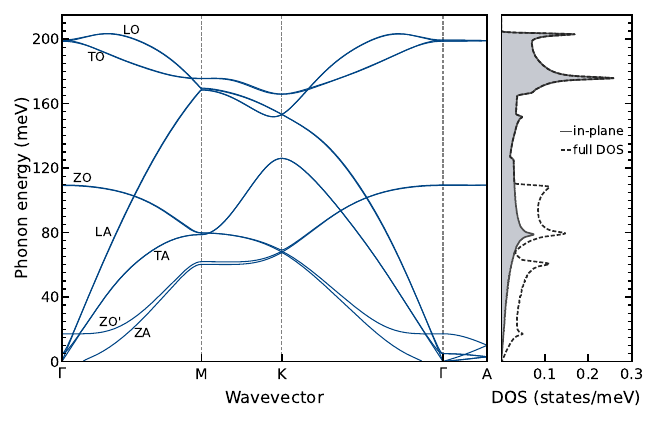}
        \caption{Left: Phonon dispersion curves of bulk graphite calculated using the r$^2$SCAN functional. Right: Corresponding phonon density of states (DOS) of bulk graphite, showing the in-plane (L$+$T) contribution (filled) together with the full phonon DOS (dashed line).}
    \label{fig:sm_graphite_dispersion}
\end{figure}

The calculated phonon dispersion of bulk graphite is shown in Fig.~\ref{fig:sm_graphite_dispersion}. The phonon spectrum comprises seven mode families, including three acoustic branches (ZA, TA, and LA) and four optical branches (ZO, ZO$'$, TO, and LO). Here, Z denotes out-of-plane polarization, whereas T and L refer to in-plane transverse and longitudinal polarizations, respectively. The corresponding phonon density of states (DOS), shown in the right panel, presents the in-plane (L$+$T) contribution as the filled area together with the total phonon DOS as a dashed line.

\end{document}